\def\year{2022}\relax
\documentclass[letterpaper]{article} 
\usepackage{aaai21}  
\usepackage{times}  
\usepackage{helvet} 
\usepackage{courier}  
\usepackage[hyphens]{url}  
\usepackage{graphicx} 
\usepackage{subfig}
\usepackage{natbib}  
\usepackage{caption} 
\usepackage{xcolor} 

\title{My Publication Title --- Multiple Authors}

\title{A Longitudinal Dataset of Twitter ISIS Users}
\author {
    Younes Karimi,
    Anna Squicciarini,
    Peter K. Forster,
    Kira M. Leavitt\\
}
\affiliations {
    College of Information Sciences and Technology,\\
    Pennsylvania State University \\
    \{younes, acs20, pkf1, kml6384\}@psu.edu
}

\begin{document}

\maketitle

\begin{abstract}

 We present a large  longitudinal dataset of tweets from two sets of users that are suspected to be affiliated with ISIS. These sets of users are identified based on a prior study and a campaign aimed at shutting down ISIS Twitter accounts. These users have engaged with known ISIS accounts at least once during 2014-2015 and are still active as of 2021. Some of them have directly supported the ISIS users and their tweets by retweeting them, and some of the users that have quoted tweets of ISIS, have uncertain connections to ISIS seed accounts. This study and the dataset represent a unique approach to analyzing ISIS data. Although much research exists on ISIS online activities, few studies have focused on individual accounts. Our approach to validating accounts as well as developing a framework for differentiating accounts’ functionality (e.g., propaganda versus operational planning) offers a foundation for future research. We perform some descriptive statistics and preliminary analyses on our collected data to provide a deeper insight and highlight the significance and practicality of such analyses. We further discuss several cross-disciplinary  potential use cases and research directions.
\end{abstract}

\section{Introduction}
 
The past few years have seen social media as an effective tool for facilitating uprisings and enticing dissent in the Middle East and beyond. \cite{lotan2011arab,starbird2012will,berger2015isis}. The embrace of social media in the Middle East has intensified the battlespace and permitted ISIS (the Islamic State of Iraq and Syria\footnote{Also known as the Islamic State (IS), the Islamic State of Iraq and the Levant (ISIL), and in Arabic, Daesh, is a militant Sunni Islamist extremist group that follows a Salafi jihadist doctrine.}) and similar groups to confront existing regimes, by spreading propaganda, recruiting sympathizers, enabling and inspiring global operations and undermining rivals \cite{berger2015isis}. 
  The FBI describes ISIS as the most adept terrorist group at using Internet and social media propaganda to recruit new members~\cite{isisonline2016}. ISIS is widely recognized for its online presence, and the importance it attributes to Twitter specifically~\cite{vidino2015isis}. With the decline of the physical caliphate, ISIS increasingly has begun to rely upon ``virtual entrepreneurs''  \cite{meleagrou2017threat}, ``fanboy''~\cite{telegram-to-twitter} and a sophisticated online media operation to propagate its goals.  
 
  Prior research efforts have attempted to capture a glimpse of these online activities,  in order to investigate ISIS online recruiting and campaigning efforts. However, to date, we are aware of very few datasets  of known extremists' online  activities \cite{kaggleisis, alfifi2018measuring}. Of these,  only a small sample is available for research  ~\cite{kaggleisis}. 

  Importantly, existing datasets refer to online activities  before 2015 which is a watershed time period. In 2015, ISIS online media efforts faced a crackdown by technology companies anxious to remove content and suspend users' accounts. ISIS had to change its online strategy. Very little is known about ISIS  online whereabouts  since then. Although most suspicious Twitter accounts are suspended or removed by now, ISIS physical caliphate has collapsed, its leader, Abu Bakr al-Baghdadi, is dead, their sympathizers and supporters are still active across a spectrum of social media applications such as WhatsApp, Telegram, and TamTam~\cite{voxpol}. Notwithstanding this transition, Twitter still provides the largest community and potential audience for ISIS supporters and a considerable number of ISIS supporters are still active on Twitter.

 In this paper, we present a large  longitudinal dataset of tweets from two sets of users that are suspected to be affiliated with ISIS, referred to as \textit{ISIS2015-21}.   
 Our collected dataset includes almost 10 million tweets of 6,173 accounts linked to ISIS and its propaganda (1,614 tweets per account on average). Importantly, these accounts' activities span from 2009 to 2021, therefore offering a longitudinal perspective to  extremists online traces.  We have found evidence of real correlations between these sympathizers and ISIS, some of which are suspended, removed, or reported as suspicious ISIS-related accounts since 2015.
 
  Our approach to validating accounts as well as developing a framework for differentiating accounts’ functionality (e.g., propaganda versus operational planning) results in a high quality dataset with unique longitudinal properties, that can serve as  foundation for future research across multiple disciplines. We perform some descriptive statistics and preliminary analyses on our collected data to provide a deeper insight and highlight the significance and practicality of such analyses. We further discuss several cross-disciplinary  potential use cases and research directions.

\section{Data Collection}

\begin{figure}[ht]
\includegraphics[width=0.47\textwidth]{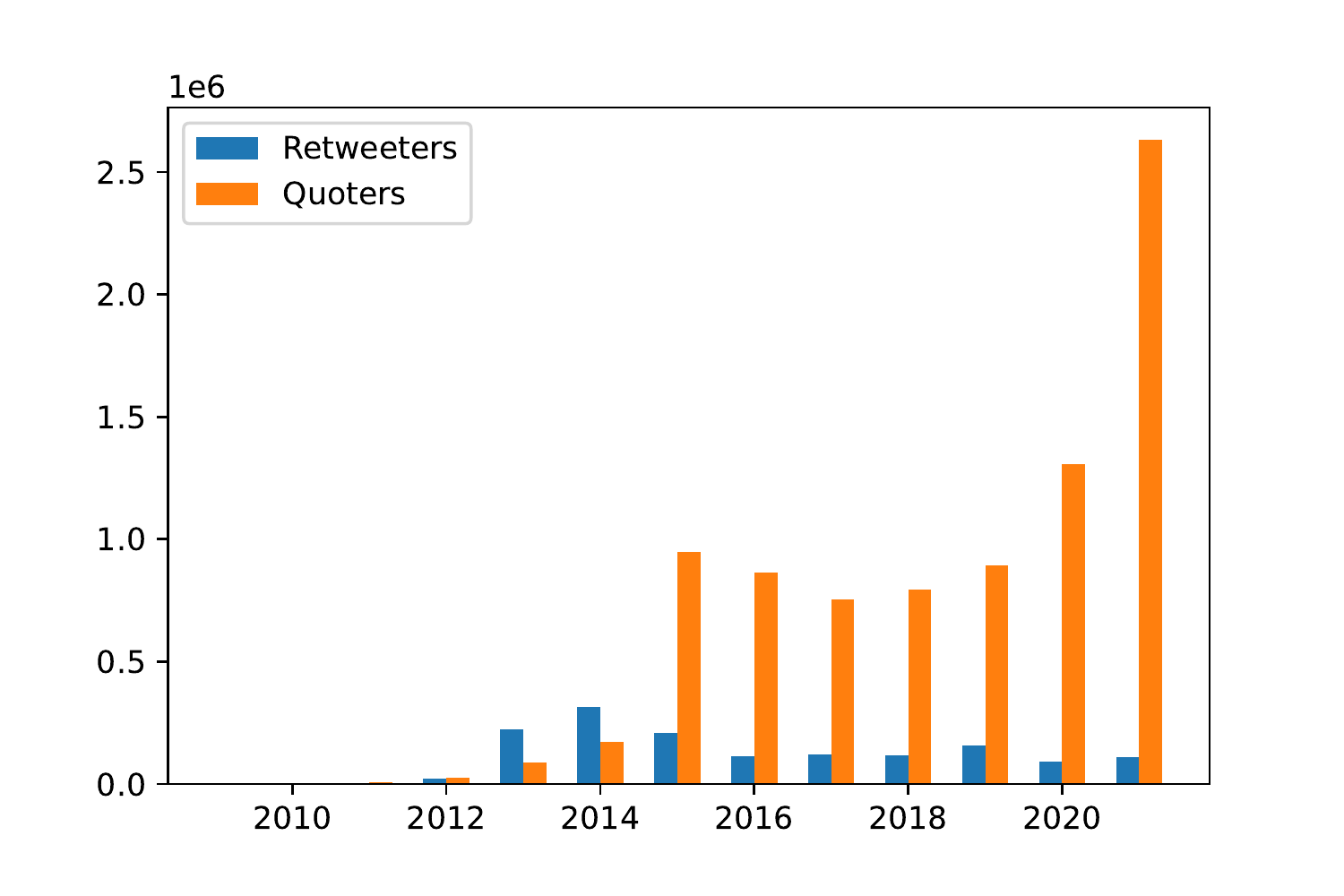}
\caption{Tweeting activity of retweeters and quoters}
\label{fig:tweeting-activity}
\end{figure}

This section reports on the temporal span and volume of the collected data, as well as analyses estimating its generalizable value. 
We used the official Twitter's API for User Tweet Timeline\footnote{https://developer.twitter.com/en/docs/twitter-api/tweets/timelines/introduction} to collect timelines from a specific set of users that are discussed below. The API enables collection of the most recent 3,200 tweets (including original tweets, retweets, replies, and quotes) that can be found on users' timelines. Collection from the timeline is limited to the most 3,200 tweets; thus, if a user has less than 3,200 tweets (all types combined) the whole timeline may be collected but if the combined tweets exceeds 3,200 only the most 3,200 are collected.  All tweets were collected during December 6–7, 2021 but depending on the account activities, they might include tweets from as early as 2009. For   users that have  more than 3,200 tweets in their timelines, most of the collected tweets  belong to 2021 (see   Figure~\ref{fig:tweeting-activity}).



\begin{table*}[ht]
\centering
\begin{tabular}{|c||c|c|c|c|c|c|c|c|c|c|} 
 \hline
 Group & Accounts & Suspended & Removed & Alive & Private & Public & Cleaned & Target & CtrlSec & Tweets\\ [0.5ex] 
 \hline\hline
  Retweeters & 1,000 & 25 & 136 & 839 & 88 & 751 & 4 & 747 & 1+ & 1,480,919 \\\hline 
 Quoters & 16,672 & 9,290 & 1,459 & 5,923 & 466 & 5,457 & 31 & 5,426 & 27+ & 8,483,234\\[0.25ex]\hline
Grand Total & 17,672 & 9,315 & 1,595 & 6,762 & 554 & 6,208 & 35 & 6,173 & 28+ & 9,964,153 \\[0.25ex]\hline
\end{tabular}
\caption{Distribution of users and collected tweets for the two groups of \textit{retweeters} and \textit{quoters} of ISIS users. All the numbers represent the corresponding number of users, except the last column which illustrates the total number of tweets. All the tweets are collected only from \textit{alive} users who have \textit{public} profiles at the time of data collection (users in the ``Target'' column). From the original set of accounts for each group, some are suspended or removed, or have made their profiles private. ``CtrlSec'' is the overlap between our users and the suspicious ISIS-affiliated users introduced by the Controlling Section Twitter accounts. Plus (`+') at the end of the numbers in ``CtrlSec'' column indicates the possibility of more overlaps, as we only looked into the accounts that were targeted in their last 3,200 tweets, while many of these accounts are alive since 2014–2015 and might have been targets of this campaign earlier.}
\label{table:users}
\end{table*}

\paragraph{Accounts description} We used an account-driven approach to generate our tweet repository and  collected about 10 million tweets generated  by 6,173 accounts during the timeline from April 2009 to December 2021 (see Figure~\ref{fig:tweeting-activity} for accounts' activities).  
We identified ISIS-related accounts  from a Twitter repository (referred to as \textit{Alfifi} dataset in what follows) originally collected by researchers during the 2014–2015 time frame~\cite{alfifi2018measuring}. In the study, authors   identified  potential ISIS-related accounts through a target crowd-sourced effort, and later studied them to detect  the markers of actual ISIS users. This crowd-sourcing effort was a campaign originally initiated by a faction of the Anonymous hacking group\footnote{https://en.wikipedia.org/wiki/Anonymous\_(hacker\_group)} called \textit{Controlling Section}\footnote{As of 2021, four Twitter users with similar usernames of \texttt{@CtrlSec} (https://twitter.com/CtrlSec), \texttt{@CtrlSec0}, \texttt{@CtrlSec1}, and \texttt{@CtrlSec2} are directly associated to this campaign.}~\cite{ctrl-sec} in 2015 that invited Arabic speakers to report Twitter accounts presumed associated  with  ISIS after several fatal ISIS attacks~\cite{anon-hacktivist}. This faction has been in charge of taking down ISIS-related websites, exposing their Twitter accounts, stealing their bitcoins, and doxing their recruiters~\cite{ctrl-sec}. The campaign is still active and exposes new suspicious and likely ISIS affiliates.  

The original Alfifi's dataset included not only tweets from presumed  ISIS  accounts, but also tweets and account information about {\em Retweeters} of ISIS—accounts that had retweeted ISIS tweets—and {\em Quoters} of ISIS—users that had quoted tweets posted by the known ISIS seed accounts at least once. In 2021,  we launched  a tweet collection of  \textit{alive} (neither suspended nor removed) and {\em public} accounts (their tweets are not protected) to collect timelines of these original accounts, so as to develop a new dataset reflecting the recent activities of potential ISIS sympathizers and ISIS affiliates that were active and engaging with known ISIS seed accounts in 2014–2015. We observed that other than the aforementioned ISIS seed accounts that were identified and suspended by Twitter around 2014–2015, the majority of  Twitter accounts in the \textit{Alfifi}'s dataset are no longer accessible and have been either suspended or removed since then. Some of these suspensions can be results of the Controlling Section's campaign, because they post suspicious user accounts and ask users to report them, so that Twitter suspends the identified malicious users.
 
In total, as reported in Table \ref{table:users}, we identified  17,672 user IDs from both \textit{Retweeters} and \textit{Quoters}. Of this users' base,  9,315 users were  suspended and 1,595 users have removed their accounts since 2014–2015 and are not included. Our data  also does not include  tweets from  a  number of   users that have protected their accounts since 2014–2015.  88 users out of the total 839 alive \textit{Retweeters} of ISIS, and 466 users out of the total of 5,923 \textit{Quoters} of ISIS that have made their profiles private. Therefore, only 6,208 accounts out of the initial 17,672 accounts were neither suspended nor removed and had public profiles at the time of our tweet collection. Additionally, 35 users had removed all the tweets from their timelines and we discarded them in our tweet collection.

Out of the remaining 6,173 users, we were able to collect  activities (i.e., up to 3,200 most recent tweets) of 4,835 users. For the remaining 1,338 \textit{Quoters}, we were only able to collect up to the 100 most recent tweets per user. This is because this API returns tweets from a user's timeline in a pagination mode, each page containing up to 100 tweets.\footnote{Even if a user has over 100 tweets in their timeline, we observed that the API may return less than 100 tweets in the first page.}  Note that we had already collected one page of tweets from all the 6,173 users before hitting the API's limit.\footnote{We initially collected one page of tweets (up to 100 tweets) from each user and then iterated through all of them to collect the rest of their timelines.Therefore, we collected more than 3,200  tweets for 25 Retweeters and 442 Quoters.} 

\begin{figure}[ht]
\includegraphics[width=0.47\textwidth]{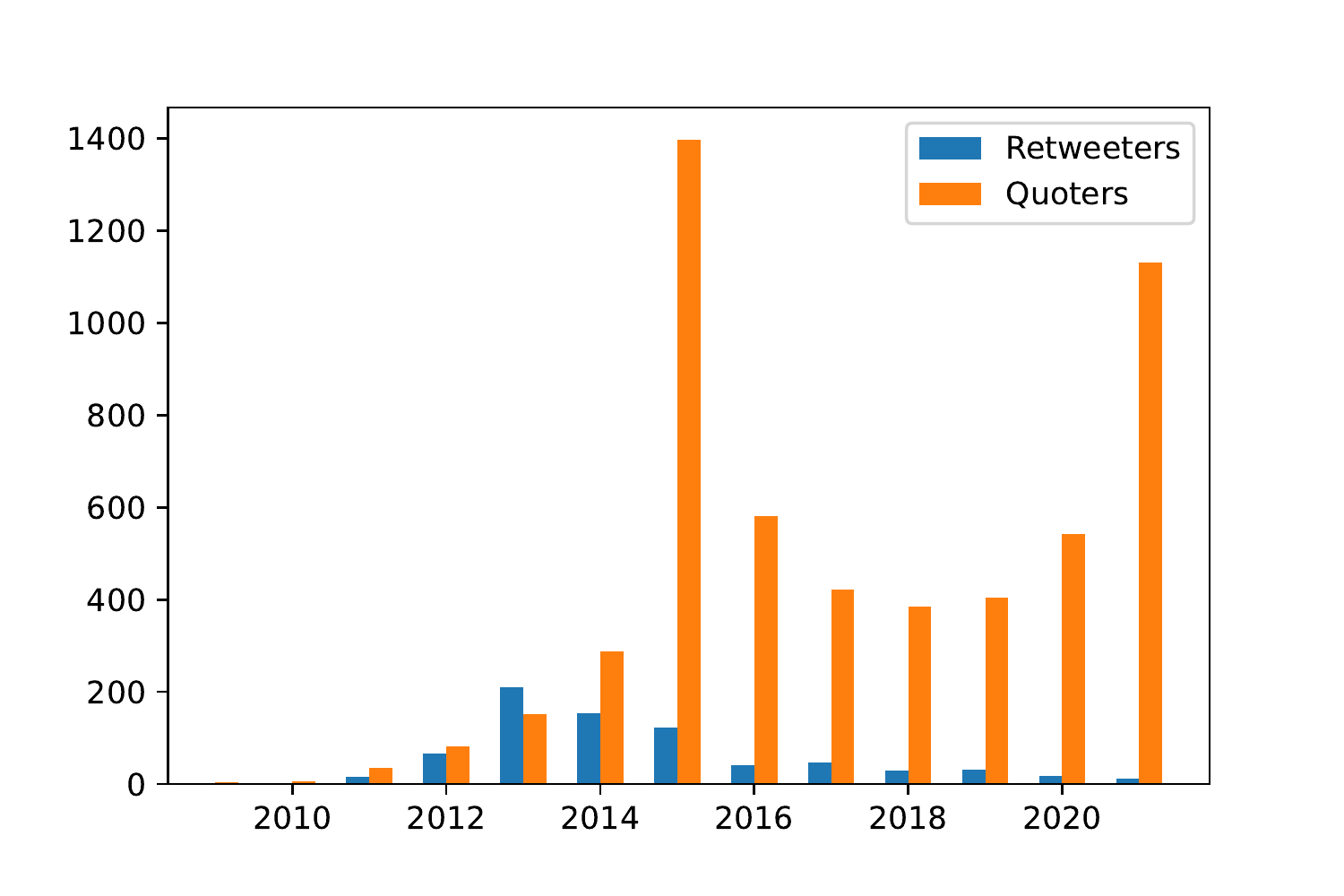}
\caption{Oldest and earliest collected tweets of the retweeter and quoter users. A considerable number of quoters have been very active in 2021 and 2015—around several ISIS attacks and large crackdown of ISIS Twitter accounts.}
\label{fig:first-activity}
\end{figure}

Figure~\ref{fig:first-activity} shows the oldest and earliest tweets that we have from these users. As it can be seen, there are two spikes in 2015 and 2021. The latter indicates that a large number of users have many tweets and have been recently active, so by collecting 3,200 tweets, we have not dug deep into their historical tweets. But, the former and our manual inspection of users shows that a considerable number of these users have been created or started their activities around the 2015 time frame when many of ISIS attacks happened.

\begin{figure}[htbp]
\includegraphics[width=0.47\textwidth]{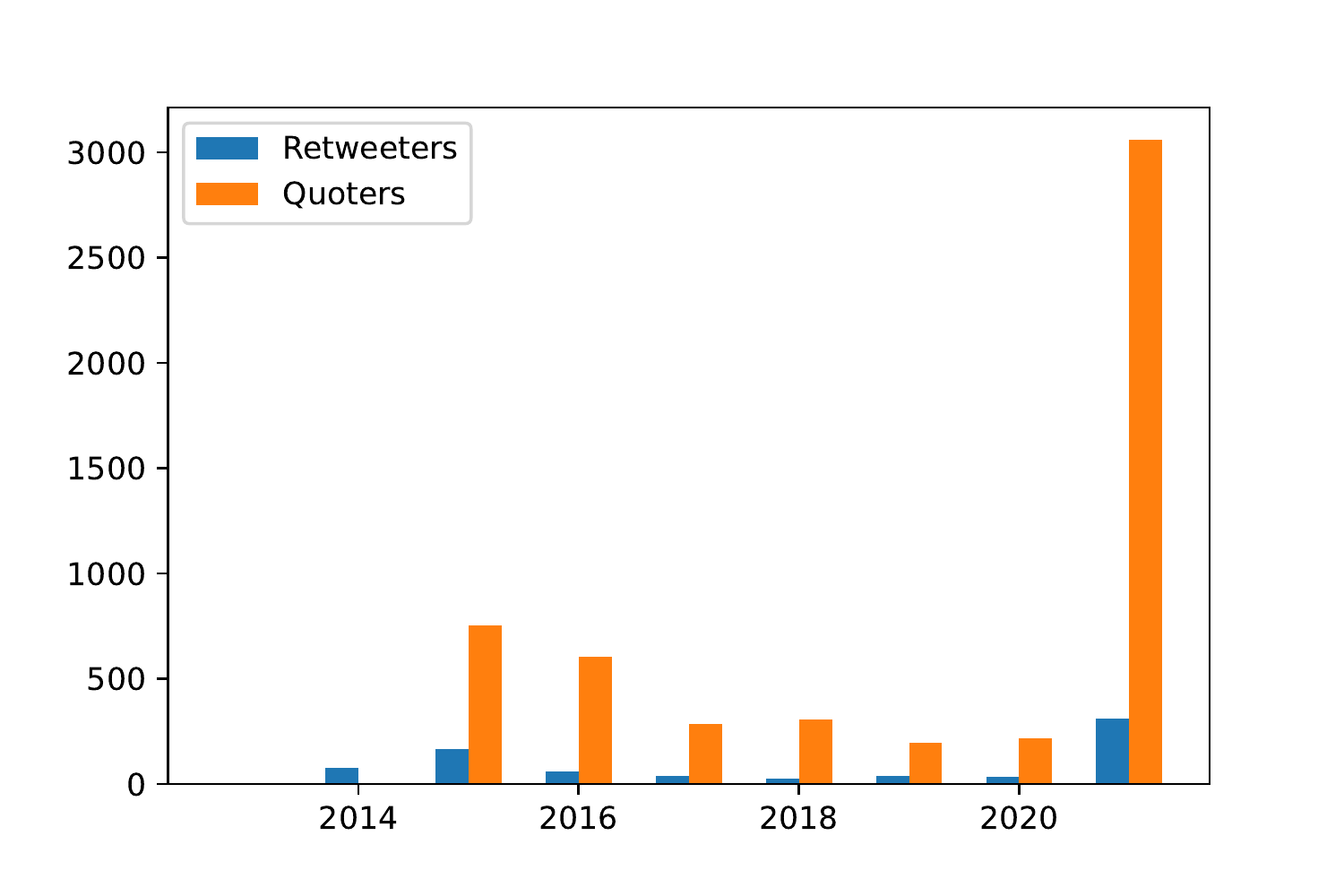}
\caption{Last and newest collected tweets of the retweeter and quoter users. A large portion of users have stayed dormant since 2015–2016.}
\label{fig:last-activity}
\end{figure}

We have also investigated \textit{dormant accounts}—accounts that have become inactive after a specific time—and have illustrated them in Figure~\ref{fig:last-activity}. As it can be observed, a considerable  number  of  users  have stayed dormant since 2015–2016—around the time the Alfifi dataset was collected and several fatal ISIS attacks had occurred.

Overall, we collected 9,840,206 tweets, from 4,835 accounts, including all the alive \textit{Retweeters} with public profiles and 4,088 \textit{Quoters} during the 2009–2021 time frame. Additionally, we had collected 123,947 tweets from the remaining 1,338 Quoters, which increased our total number of tweets to 9,964,153. None of the original ISIS accounts identified by Alififi \cite{alfifi2018measuring} was found alive in December 2021. Furthermore, out of 4,835 users for whom we collected the whole timelines, two Retweeters and 403 Quoters had less than 3,200 historical tweets in their timelines and the rest of accounts had more activities.

In Table~\ref{table:users}, we also report data for the {\em Cleaned}  accounts, i.e. accounts  that  have deleted all the tweets from their timelines.  We have discarded the cleaned accounts from our final number of users that are shown in the ``Target'' column.

\paragraph{Accounts' validation} 
 
The tweets in our dataset are likely from ISIS sympathizers, but the degree of their involvement is not certain.
Retweeting can be considered as an act of support and help for spreading a message to a larger audience. It has been shown ~\cite{alfifi2018measuring} that retweeters of ISIS are more likely to engage in malicious activities. Additionally, we may observe quotes of other tweets when someone tries to either support or oppose an ISIS-related message or user. In addition to the previous suspicious activities and potential connections to ISIS that we have observed from these users in the old dataset, we have looked up their user IDs in the Controlling Section's target suspicious accounts that is a more recent list of potentially ISIS-related users. 

It is extremely challenging to fully ascertain an account's origin and its actual affiliation. 
Yet, we have strong indicators that our dataset includes several  accounts are not only ISIS-related but most likely, are from ISIS. We found some overlap among the accounts we collected and the accounts reported as ISIS (see Table~\ref{table:users}), under the ``CtrlSec'' column. This highlights that while some of these suspicious accounts could have been identified and suspended around the time that \textit{Alfili}'s paper was published, they were not shut down by Twitter yet and had remained alive for a while. We also observed three new suspensions and 13 users that removed their accounts during our two-day data collection which is indicative of  the suspicious nature of our target users.

 

\begin{table*}[!htbp]
\centering
\begin{tabular}{|c||c|c|c|c|c|} 
 \hline
 Group & Original Tweet & Retweet & Quote & Reply & Quote \& Reply \\ [0.5ex] 
 \hline\hline
  \rule{0pt}{1ex} Retweeters & 765,128 & 513,087 & 10,677 & 192,375 & 348\\\hline 
 Quoters & 2,176,986 & 3,678,238 & 443,657 & 2,200,187 & 15,834\\[0.25ex]\hline
\end{tabular}
\caption{Activity breakdown for the tweets. Original tweets are those that are not retweeted, quoted or replied to another tweet. The last column shows the number of overlapped (duplicate) tweets amongst the two previous columns.}
\label{table:activity-type}
\end{table*}

\begin{table*}[!h]
\centering
\begin{tabular}{|l||l|l|c||l|l|} 
 \hline
 Group & Top Hashtags & Count & Total Hashtags & Top Keywords & Count\\ [0.5ex] 
 \hline\hline
  \rule{0pt}{1ex} & tweet in remembrance of Allah* & 97,492 && Allah* & 340,281\\
  & muslim's treasure* & 43,971 && o Allah* & 245,484\\
  & prayer to my lord* & 27,494 && martyred* & 74,663 \\
  & Quran & 19,195 && said* & 49,564\\ 
  Retweeters & hadith & 16,899 & 599,121 & face with tears of joy emoji & 41,812 \\
  & Allah & 8,484 && god* & 38,359 \\
  & publish his biography* & 6,298 && peace be upon him* & 38,352 \\
  & Saudi Arabia* & 4,511 && lord* & 33,822 \\
  & tweet hadith* & 4,269 && peace* & 32,266 \\
  & crescent moon* & 3,241 && day* & 31,973 \\\hline 
  
 & Saudi Arabia* & 53,477 && Allah* & 1,312,104 \\
 & Iraq* & 37,011 && face with tears of joy emoji & 534,122\\
 & Syria* & 36,063 && o Allah* &  404,598 \\
 & Immediate* & 28,631 && said* & 178,785\\
 Quoters & Iran* & 26,680 & 3211867 & today* & 171,952\\
 & Turkey* & 25,161 && swear to Allah* & 152,045\\
 & Qatar* & 23,239 && white heart emoji & 144,907\\
 & Yeman* & 21,329 && Muhammad* & 136,848\\
 & Egypt* & 21,218 && the people* & 130592 \\
 & Quran & 16,580 && day* & 128,641\\
 [0.25ex]\hline
\end{tabular}
\caption{Top 10 most frequent words in tweet texts. Phrases with an asterisk (*) are translated. For the phrases that were originally written in English, we have reported the aggregate count disregarding their casing.}
\label{table:top-hashtags-keywords}
\end{table*}


\begin{table}[!htbp]
\centering
\subfloat[Retweeters]{
\begin{tabular}{|l||l|} 
 \hline
 Language & \# Tweets\\ [0.5ex] 
 \hline\hline
  \rule{0pt}{1ex}
  Arabic & 1,384,028\\\hline
  Undefined & 51,645\\\hline
  English & 37,625\\\hline
  Indonesian & 2,057\\\hline
  Spanish & 491\\\hline
  Korean & 486\\\hline
  Catalan & 424\\\hline
  Persian & 369\\\hline
  French & 350\\\hline
  Tagalog & 341\\\hline
  Other & 3,103\\[0.25ex]\hline
\end{tabular}}
\quad
\subfloat[Quoters]{
\begin{tabular}{|l||l|}
 \hline
  Language & \# Tweets \\ [0.5ex] 
 \hline\hline
  \rule{0pt}{1ex}
  Arabic & 7,632,693\\\hline
  Undefined & 482,756\\\hline
  English & 297,003\\\hline
  French & 18,819\\\hline
  Persian & 9,460\\\hline
  Indonesian & 6,926\\\hline
  Turkish & 6,205\\\hline
  Spanish & 4,780\\\hline
  Hebrew & 3,252\\\hline
  Catalan & 3,090\\\hline
  Other & 18,250\\[0.25ex]\hline
\end{tabular}}
\caption{Distributions of top-10 frequent languages used in our tweets' texts. ``Other'' includes 38 and 49 other languages that were identified in our collection with lower frequencies for Retweeters and Quoters respectively. Numbers represent the number of tweets that are written in the corresponding languages.}
\label{table:labguages}
\end{table}

\section{Data Sharing and Format}
Our ISIS2015-21 dataset is available for download under FAIR principles~\cite{wilkinson2016fair} in CSV and ZIP formats. The data includes two tables for retweeters and quoters, and tweets are keyed by Twitter assigned IDs and augmented with additional metadata as described below. Moreover, the data includes unique keys for different types of media that appeared in the tweets, including photos, videos, and animated GIFs, as well as aggregated tables for the websites and URL domains appeared in the tweets of each the two groups along with their corresponding frequencies.

The  ISIS2015-21  dataset conforms with FAIR principles.\footnote{https://www.force11.org/group/fairgroup/fairprinciples} The dataset is \textit{Findable} as it is publicly available on Harvard Dataverse.
It is also \textit{Accessible} because it can be accessed by anyone around the world.
The original dataset is in CSV format, and contains all the tweet IDs that can be utilized to lookup or download the tweets, therefore it is Interoperable. We release all the tweet IDs and user IDs in our dataset with descriptions detailed in this paper
making the dataset \textit{Reusable} to the research community.

The dataset  also includes a list of  user IDs of suspended and removed accounts, for potential archival research. These tables do not include raw tweet data beyond the ID, according to Twitter’s Terms of Service. However, to support use of the data without being required to download (``hydrate'') the full set of tweets, we augment the Tweets table with several key properties. For each tweet
we provide the author's ID, the conversion ID that the tweet belongs to, the timestamp for the tweet, the number of total retweets, replies, quotes, and likes as computed by Twitter. The Tweets table properties also include the language of the tweet as identified by Twitter, whether the tweet contain sensitive and Not-Suitable-For-Work (NSFW) content, the source or the app used to post the tweet, and the IDs of the retweeted, replied, or quoted tweets where they apply (ID of zero elsewhere).

\section{Use Cases}
The primary motivation in creation of this dataset is to study
 ISIS's evolving presence  on social media in the last 10 years. Following is a series of use cases that permit the analysis of ISIS methods for disseminating propaganda or the "tools of the trade"; improving message longevity and amplification increasing the possibility of resonance within target populations, and ultimately creating the link between the individual and group. We also
provide use cases with a non-ISIS focus such as studying conversations and hate speech. Note that this list is not
exhaustive.
\subsection{Language and Network Analysis}
An expected contribution to this dataset is to foster researchers' understanding of language and network patterns  of extremists online. Data-driven studies in this space have highlighted some interesting findings so far, such as the limited social network reach of ISIS sympathizers and a self-reinforced network of users~\cite{berger2015isis}. Yet, several of these findings have not been validated on online activities post 2015 in the wake of tech companies crackdown and the disintegration of physical caliphate which provided some resources to the media operations. We anticipate that in light of  these events and the new attentive eye of Twitter toward extremists' activities,  ISIS members and sympathizers have morphed their online  behavior to evade detection and yet still spread new messages of relevance to them. 
This new dataset can be used to extract recent  online patterns of behavior, online signatures, and online resilience. Furthermore,
one could  examine the evolution of both ``successful'' and ``unsuccessful'' ISIS sympathizers in our dataset,  to identify the impact of existing interventions already present the dataset. 

The new Twitter's API (API v.2) also identifies and provides a set of annotations for each tweet containing some of its named entities such as \textit{Person} names, \textit{Places}, \textit{Organizations}, \textit{Products}, and \textit{Others} that do not fill into these categories.\footnote{https://developer.twitter.com/en/docs/twitter-api/annotations/overview} Table~\ref{table:annotations} illustrates the frequency of each of these named entities in our dataset as well as the number of hashtags for each group of users. 

\begin{table*}[!htbp]
\centering
\begin{tabular}{|c||c|c|c|c|c|} 
 \hline
 Group & Person & Place & Organization & Product & Other\\ [0.5ex] 
 \hline\hline
  \rule{0pt}{1ex} Retweeters & 119,756 & 78,662 & 33,093 & 7,670 & 3,259\\\hline 
 Quoters & 1,040,521 & 1,508,607 & 265,898 & 56,554 & 10,506\\[0.25ex]\hline
\end{tabular}
\caption{Named entities and hashtags used in the dataset  tweets.}
\label{table:annotations}
\end{table*}

As this metamorphosis evolves, we see potential in  investigating individual-level models of influence in domains where information sharing is driven by political and extremist agenda and the receiving population may be highly heterogeneous.  One approach might be to  focus on creating a network-based   behavioral model, as well as  an  individual-level  model that can capture imitative behaviors. This analysis builds on earlier insights into how extremists imitate each other through songs, martyr posters, and pledges of allegiance.One tool for analysis is the study  of   discourse-level contagiousness that drives extremism to uncover overall patterns of extremist ideology infectiousness.  ISIS uses multiple platforms and information tools to promote its cause, influence behavior, and conduct operational business. The process is easy to understand. Create the desired message and disseminate it to either a targeted or broad audience depending upon the goal. The objective is to have the message “stick”  or shape a conversation around a topic thus influencing thought and/or behavior. 

More specifically,  contagious online discourses consist of a series of online messages that agree with the others in certain subjectivity dimensions, such as sentiment polarity (e.g., unanimously positive or negative) or emotion types (e.g., full of anger), or in writing styles (e.g., exaggeration using similes, slang, or punctuation).  
Negative online contents are often inflammatory or extremely emotional.   We can envision using  sentiment and emotion analysis methods such as lexicon based sentiment classification~\cite{jurek2015improved}, to develop tools to analyze rich extra-propositional properties of user generated texts in order to facilitate modeling contagiousness of online discourses.

\subsection{Analysis of ISIS Propaganda}
 Beyond contagion, we can investigate  propaganda messages in our dataset. A propaganda campaign   is a combination of formal structure and bottom-up information dissemination coordinated by the Diwan of Central Media. ISIS’s presence combines the formal ``news agency,'' AMAQ, with direct messaging applications (e.g., Twitter, Telegram, or YouTube), and a Dark Web presence. Understanding the overall media structure is helpful. However, this structure introduces a challenge in that each of these environments may be used for multiple tasks complicating the identification of relevant propaganda pieces.  Thus, a relevant analytic model needs a data fusion approach. For the purposes of this study which focuses exclusively on Twitter, three processes provide a basis:
\begin{itemize}
    \item 
	Identify relevant propaganda pieces by understanding that propaganda and operational content differ,
	\item	Recognize indicators of message distribution aimed at increasing sharing which reflects propaganda, 
\item	Explore message content in terms of accuracy, media used and language.
\end{itemize}

According to the US government, terrorist propaganda provides pervasive and continuous condemnation of the ``other side,'' and proposes ideological alternatives which they know would not be generally acceptable if they were openly revealed~\cite{international1975clandestine}.  Two components from this definition help to broadly identify a method for understanding terrorist propaganda and can be instrumental for detection and analysis in our dataset.    To be ``pervasive and continuous'' message sharing is critical. Whereas an operational message has a specific audience, a propaganda message demands broad dissemination. Thus, through information and network analysis, messages in which (1) the creator has few followers but (2) whose message is widely disseminated via retweets among a larger network likely indicates propaganda. For instance, we identified a relatively highly disseminated tweet\footnote{The IDs of the tweet and the user were 1182737917120307201 and 965077656 respectively.} stating religious content that only had 108 followers at the time of data collection, but the tweet had gained 305 retweets, 327 likes, and 18 replies, which are clearly indicative of propaganda. 

\begin{figure}[ht]
\includegraphics[width=0.47\textwidth]{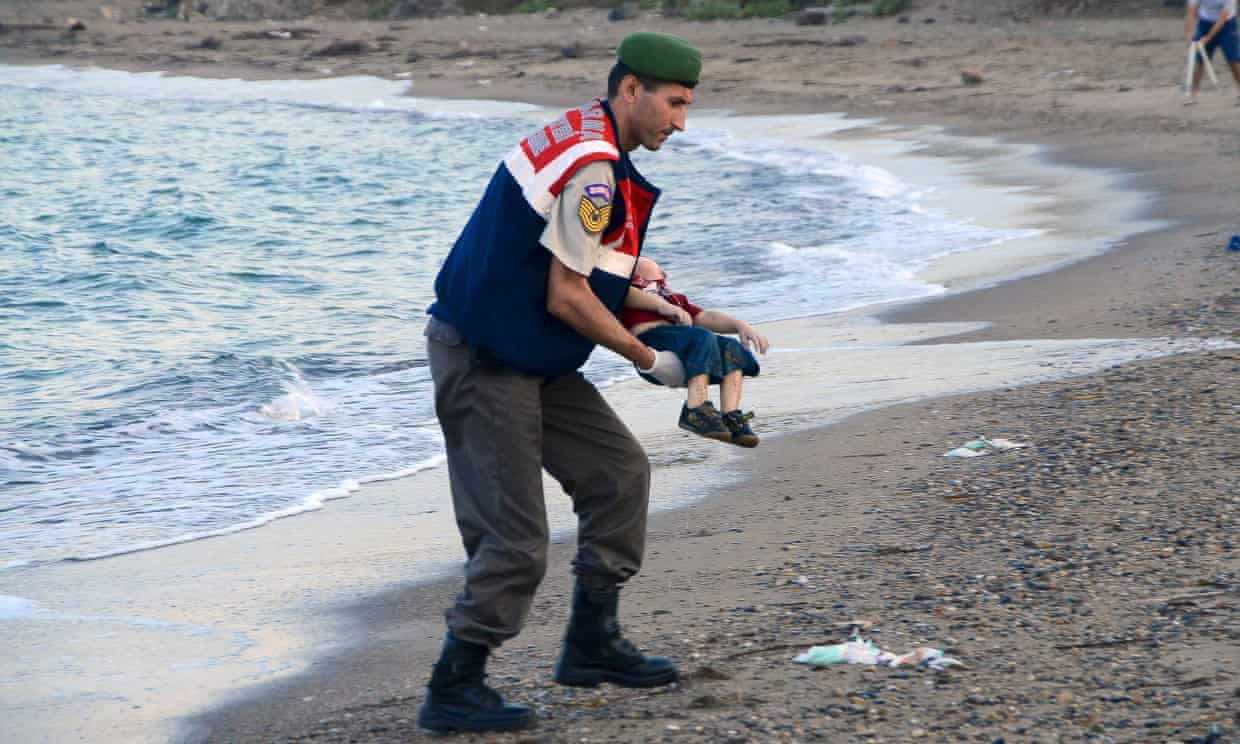}
\caption{ISIS re-published photo in social media for propaganda purposes}
\label{fig:drowned-syrian-boy}
\end{figure}

Second, propaganda is seeded with visceral and ideological based verbiage and images. These become important in identifying something as propaganda. Whereas, operational messages deal with targeting, methods, and timing and might include images (e.g., an aerial picture of a targeted building) that have tactical importance, the language and images in propaganda messages seek to elicit an emotional response. For example, ISIS re-published the photo in Figure~\ref{fig:drowned-syrian-boy}, originally published in The Guardian, with an article citing the consequences of leaving the caliphate in its English-language magazine, Dabiq~\cite{theguardian}.

A third indicator of ISIS propaganda is the presence of \textit{hashtags}. In an effort to enhance the echo chamber to which social media users gravitate, ISIS organized a sophisticated hashtag campaign. They recruited people to retweet hashtags to create trending ideas, hijack trending hashtags, curate group messaging to improve branding, and ensure message longevity even if the original platform removes it~\cite{telegram-to-twitter}. Table~\ref{table:top-hashtags-keywords} lists some of most frequent hashtags that are mostly about strong religious references amongst the retweeters, and amongst the quoters, most of them are clearly referring to middle-east countries. Manual inspection shows that many of hashtags with middle-east country references are news-related.

\subsection{Governance and Content Moderation Monitoring}
Although Twitter abrogates some responsibility for ``sensitive'' content to its users, asking them to self-select, it has established policies around graphic violence, adult content, and hateful imagery.\footnote{https://help.twitter.com/en/rules-and-policies/media-policy} Violations may be reported by Twitter users for review or identified by Twitter's algorithm. If determined offensive, Twitter takes enforcement actions either on a specific piece of content (e.g., an individual Tweet or Direct Message) or on an account if continuous violations are identified. It may also employ a combination of these options if the behavior violates the Twitter Rules.\footnote{https://help.twitter.com/en/rules-and-policies/enforcement-options} Notwithstanding, controversy exists regarding the effectiveness of the algorithm and the extent to which it is used.\footnote{https://www.quora.com/Does-Twitter-allow-adult-content} Within the terrorism environment, some things  such as a video of a beheading are easily identified and removed. Additionally accounts associated with terrorist organizations or individuals may be suspended \cite{berger2018extremism, isisonline2016}. We saw this occurs within the current dataset. However, some content evades the algorithm and may not be recognized as terrorism propaganda by users and thus continues of exist. Furthermore, user adaptability, persistence, and technical skills in hyperlinking content continue to circumvent controls. 


\subsection{ISIS Media and Redirection Strategies}

\begin{figure}[ht]
\includegraphics[width=0.47\textwidth]{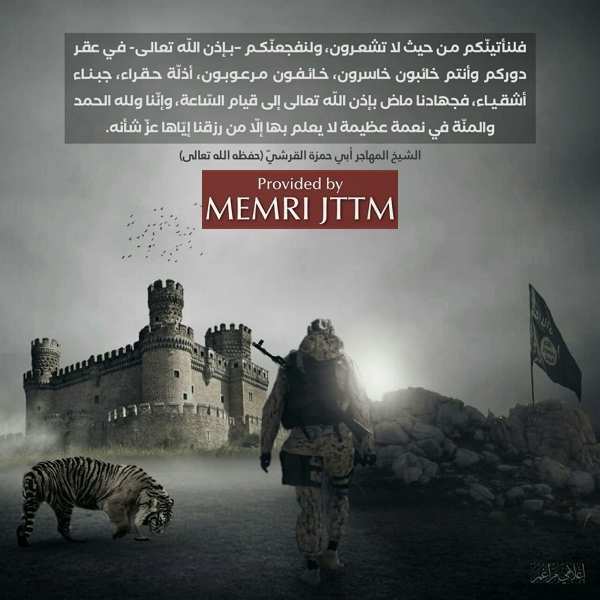}
\caption{An ISIS poster titled as ``We Will Come To You From Where You Do Not Expect'' aimed at establishing fear for the enemy while inspiring action.}
\label{fig:memri-poster}
\end{figure}

As previously noted, ISIS has a diverse online media network that facilitates the dissemination of information through multiple methods and forms of content with varying sophistication. Additionally, ISIS has become masterful at redirecting users to both enhance operational security and improve message longevity. Pictures, posters, memes, and language are all tools of the ISIS propaganda machine. In December 2021, a new ISIS group, I'lami Muragham, posted four posters on Telegram proclaiming an ISIS jihad to the ``last day'' and encouraging resistance~\cite{isismedia}. The poster in Figure~\ref{fig:memri-poster}, entitled, ``We Will Come To You From Where You Do Not Expect,'' is aimed at establishing fear for the enemy while inspiring action. The message is a quote from ISIS spokesman, Al-Shaykh Al-Muhajir. 

A second poster in the same release contained a quote by the dead leader al-Baghdadi. As a result, research into developing processes for identifying the presence of posters along with a quotation from a leader process may help classify messages as propaganda. 

ISIS' media strategy shows its sophistication by reiterating messaging and adaptability to increase messaging longevity. The former is accomplished through message re-enforcement, for example, repeating messages in various environments. A poster distributed via Telegram references a quotation from a specific issue of al Naba, ISIS' weekly newspaper. As a result, the messages are mutually reinforcing as well as offering the receiver links to other forms of propaganda that further promulgate the echo chamber increasing the likelihood resonance. Another strategy are adaptable measures meant to protect communications. Previous research contends that tabulating suspended accounts as a  measure of counter-terrorism policy is incomplete and deceiving~\cite{weirman2020hyperlinked}.
This is significant for the research on propaganda because it indicates that the suspension of accounts\textit{ does not} eliminate the propaganda. Strategies include active measures to preserve communications when accounts are suspended and to re-direct users to newly created accounts or a return to a particular platform.

An active measure to insulate networks is through file-sharing of URLs (Ibid). In our  dataset, we have identified a large set of URLs and media content that should be analyzed. The presence of URLs or truncated URLs may well be indicative of propaganda but also provide a mechanism to further exploit the information's path in an effort to suspend future versions of the message.   In Table~\ref{table:urls} we report some examples of URL domains that we found in our dataset. As shown, several of these URLs are related to other potential propaganda outlets (e.g., YouTube) and religious websites (e.g., ``tweet4allah.com'').  

\begin{table}[!htbp]
\centering
\begin{tabular}{|c||c|c|c|c|} 
 \hline
 Group & Image & Video & GIF & URL \\ [0.5ex] 
 \hline\hline
  \rule{0pt}{1ex} Retweeters & 121,788 & 32,007 & 3,485 & 289,180 \\\hline 
 Quoters & 999,985 & 317,134 & 24,738 & 2,140,794 \\[0.25ex]\hline
\end{tabular}
\caption{Distribution of different types of media and URLs that are uniquely extracted from the collected dataset.}
\label{table:media-types}
\end{table}

\begin{table}[!h]
\centering
\begin{tabular}{|l||l|} 
 \hline
 Retweeters & Quoters \\ [0.5ex] 
 \hline\hline
  du3a.org & du3a.org\\
  knz.tv & youtu.be\\
  tweet4allah.com & fb.me\\
  gharedly.com & d3waapp.org\\
  quran.ksu.edu.sa & www.facebook.com\\
  knz.so & ghared.com\\
  d3waapp.org & quran.to\\
  3waji.com & www.youtube.com\\
  knzmuslim.com & bit.ly\\
  ghared.com & ln.is\\
  [0.25ex]\hline
\end{tabular}
\caption{Most frequent domains amongst retweeters and quoters. While there are shared domains amongst the two groups, retweeters have higher uses of Islamic apps and URLs and quoters have higher uses of redirection to other media sharing services. We have omitted ``twitter.com'' which stands for the most frequent domain for both groups.}
\label{table:urls}
\end{table}

\begin{table}[htbp]
\centering
\begin{tabular}{|l||l|} 
 \hline
 Retweeters & Quoters \\ [0.5ex] 
 \hline\hline
  Twitter for iPhone & Twitter for iPhone\\
  Twitter for Android & Twitter for Android \\
  Muslim Treasure* & Twitter Web Client \\
  Quranic App* & Twitter Web App\\
  Quran Tweets* & Twitter for iPad\\
  GharedlyCom & Facebook\\
  Twitter for BlackBerry\textsuperscript{\textregistered} & Quranic App*\\
  Twitter for iPad & Quran Tweets*\\
  Twitter Web Client & Verse App*\\
  Ayat & Tweetbot for iOS\\[0.25ex]\hline
\end{tabular}
\caption{Top sources of tweets and applications used to post tweets. Asterisk (*) indicates that we have translated the original Islamic application names from Arabic.}
\label{table:sources}
\end{table}

\subsection{Data-Informed Terrorism Studies} 
Political terrorism relies on information operations to amplify an incident, create fear beyond the event itself, establish the group's unity, identity, and ideological foundation creating an ``ingroup'' versus ``outgroup'' dynamic~\cite{berger2018extremism} and/or provide operational direction. Within this environment, propaganda fills a vacuum created by the lack of power. It may take the form of being event-driven seeking to exploit the incident by having it trend among a broad audience or more formally it is the release of information via high quality videos, magazines, or newsletters. 

ISIS and its affiliates (e.g., Islamic State – Khorasan Province (ISIS-K, IS-K or IS-KP)) have a specific process for achieving their goals. First, they seek to promote chaos both physically and virtually to demonstrate the inability of an existing entity to protect its people. From a propaganda perspective this might include videos, pictures, and comments promoting activities (i.e., violence) to demonstrate ineffectiveness.  Second, as previously discussed, it seeks contagion to draw others to the cause. ~ 

A recent post on ISIS-operated Rocket.Chat  
encourages long-wolfs in the West provides an example of the latter:
``Attack the citizens of crusader coalition countries with your knives, 
run them over in the streets, detonate bombs on them, and spray them 
with bullets. Those who preceded you on this path, who frightened the 
Crusaders and robbed them of safety and security in their own homes, 
like their deeds, may Allah accept them and replicate them ...''~\cite{JihadisChristmas}.

Operationally, ISIS22015-2021 dataset can be used to analyze the online activity (e.g. messaging, comments, and propaganda posts) surrounding successful real world acts of terrorism from of IS-K and its affiliates. Such activity is important both from a validation and analytic perspective. Evaluating messaging linked to real world events further strengthens the identification of propaganda and perhaps even allows a categorization of types or levels of propaganda. By comparing and contrasting Twitter commentary with reliable sources, such as multiple press comments surrounding an event, a presumption of a tweet being propaganda may be made and even measured based upon the amount of true versus fabricated information presented.  This process requires further refinement. 

\begin{figure}[ht]
\includegraphics[width=0.48\textwidth]{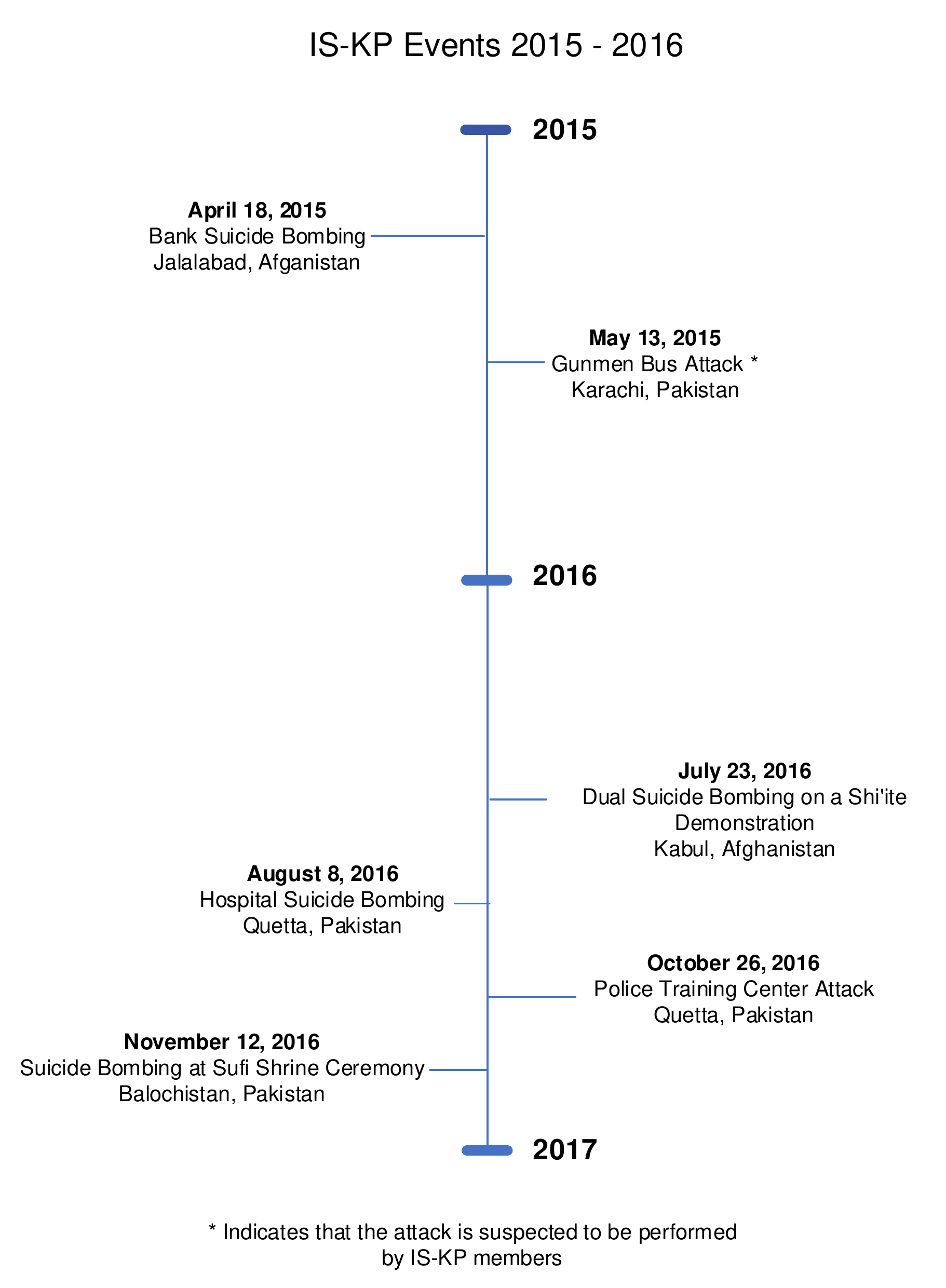}
\caption{Timeline of ISIS attacks during 2015–2016}
\label{fig:Timeline2015-2016}
\end{figure}

\begin{figure}[ht]
\includegraphics[width=0.48\textwidth]{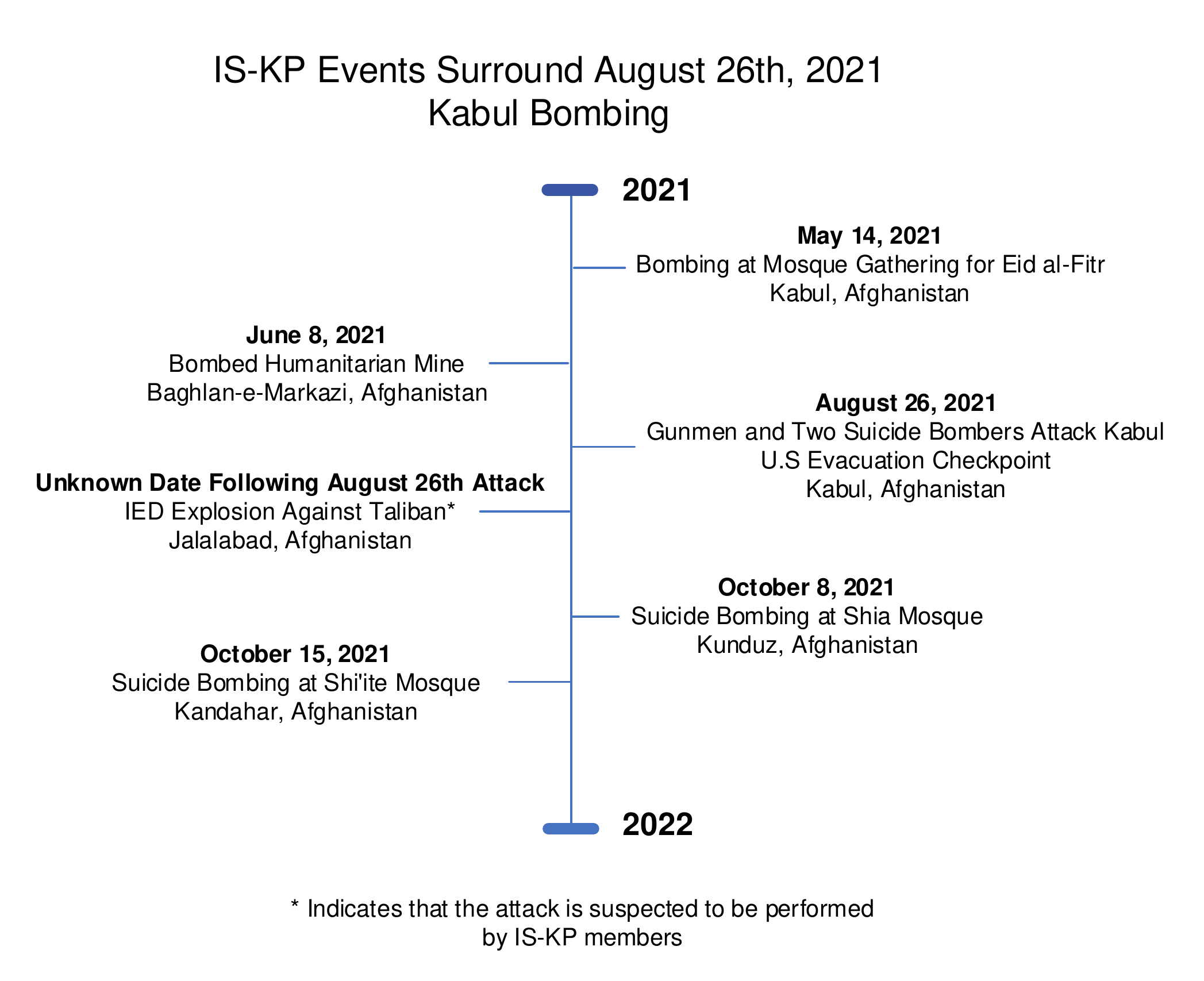}
\caption{Timeline of ISIS attacks during 2021}
\label{fig:Timeline2021}
\end{figure}

Since terrorism’s goal is to establish fear, a terrorist group can use online activity to connect their beliefs and propaganda by exploiting a successful terrorist attacks and media attention. The timelines in Figure~\ref{fig:Timeline2015-2016} and Figure~\ref{fig:Timeline2021} show IS-K events, which have the potential to coincide with an increase in online activity of terrorist groups. As it was shown in Figure~\ref{fig:first-activity}, the first activity of many quoters in our dataset goes back to 2015, when IS-K was formed.\footnote{https://cisac.fsi.stanford.edu/mappingmilitants/profiles/islamic-state-khorasan-province} We have also searched for the presence of a set of keywords related to the attacks depicted in the timeline of Figure~\ref{fig:Timeline2021}. The frequency of each of these keywords for each of the two groups of users are reported in Table~\ref{table:attacks}. Further analysis can  study  the contexts of these keywords within tweets and uncover their correlations with IS-K attacks.

The online activity (reflected in ISIS2015-21)  will likely be localized to the region most impacted by the attack. It can be speculated that online activity from international regions about an attack indicates a higher severity. The large audience generated by international communications will allow for a wider spread of propaganda and benefit the basic terrorism goal of contagion to attract new members. Traces of ISIS2015-21 can be instrumental to the analysis of such propaganda.  In particular, accounts means and frequency of communication  provide information about a terrorist group's developmental stage. For example, a two-level propaganda strategy in which a broad-based of popular support is built from which to inspire violent attacks is typical of more mature terrorist organizations. In addition, internal propaganda (written only for members of the terrorist group) presages the external propaganda written for the public and provides guidelines on the part violence is to play~\cite{international1975clandestine}. Furthermore, the messaging process reflects maturity as well. Continuous messaging requires sustainability in terms of new content, message longevity, ad platform resilience which is difficult to achieve without a robust structure. 


\begin{table}[!htbp]
\centering
\begin{tabular}{|l||l|l|} 
 \hline
 Keyword & Retweeters & Quoters\\ [0.5ex] 
 \hline\hline
 mosque & 4,989 & 36,881\\\hline
 al-fitr & 3,095 & 15,151\\\hline
  bomb & 633 & 9,541\\\hline
  gun & 423 & 4,653\\\hline
  suicide & 316 & 5,204\\\hline
  taliban & 257 & 17,462\\\hline
  kabul & 61 & 2,240\\\hline
  kandahar & 10 & 489\\\hline
  jalalabad & 1 & 46\\\hline
  kunduz & 1 & 30\\\hline
  baghlan & 0 & 107\\[0.25ex]\hline
\end{tabular}
\caption{Frequencies of 11 keywords related to IS-K attacks in the dataset. We have combined the frequencies for each term in English and Arabic.}
\label{table:attacks}
\end{table}

 \subsection{Beyond the Data: Nudging Anti-Normative Behavior}
In addition to exploratory research, this dataset can be used as a seed  to design and validate new interventions lessening the impact of extremist group influence. Upon  examining the evolution of ISIS sympathizers in our dataset,  we may  identify the impact of existing interventions already present in the dataset (e.g., account suspension or tweet removal), and then design new anti-normative nudges based on generative language models that are informed by learned models of extremist influence. 
While there is evidence that networks and beliefs collectively evolve~\cite{lazer2010coevolution}, interventions designed to shape this evolution must be carefully designed to avoid unintended effects  \cite{Bail9216,nyhan2010corrections}.  For example, an anti-ISIS message  inserted into the network to maximize exposure to ISIS sympathizers could backfire, instead strengthening the ISIS support of the sympathizers. Accordingly, carefully designed solutions must be engineered for effective  anti-normative mechanisms, building on recent advances in deep learning for generating natural language~\cite{DBLP:journals/corr/BowmanVVDJB15,DBLP:journals/corr/abs-1708-02709}.

\section{Limitations}
Our proposed ISIS20215-21 dataset  is innovative and unique in its kind, due to the specific sub-population it accurately targets and its evident longitudinal properties. Given how accounts were originally selected and their validation obtained through manual inspection and cross-correlation with CtrlSec data, we are confident about the authenticity of the accounts and their relevance for online extremist research.
Nevertheless, our dataset suffers from some limitations that should be aware to researchers interested in the data's future use.  First,  due to the existing limitation in Twitter's API, we cannot collect the complete history of each account. This however, still allows us  to include tweets for many accounts that span few years of online presence (see Figure \ref{fig:first-activity}).

Second,  even though the dataset offers a good excerpt of ISIS' evolution, there is a growing evidence that ISIS may have moved away or shifted  from Twitter for most of its online activities~\cite{voxpol}. Accordingly, while we can get a look at sympathizers and their messages, we may be missing a significant portion of their online presence and activities. Related, we suspect that there may be  other  accounts, that we \textit{do not }include in our dataset, that are close to ISIS.  We are also unable to potentially remove noisy tweets, irrelevant to our study. However, results of our preliminary analysis on the data strongly suggest that our collected content is relatively clean, and  will constitute a strong  seed    for researchers in this space.


\section{Conclusion}

 In this paper, we present a new public dataset tracking
  tweets from two sets of users that are suspected to be ISIS affiliates. These users are identified based on a prior study and a campaign aimed at shutting down ISIS Twitter accounts.
 This study and the dataset represent a unique approach to analyze ISIS data. Although some research exists on ISIS online activities, few studies have focused on individual accounts. 

The long-term aim of this project is to tackle the ambitious challenge of linking social media observations directly
to online political extremism. We hope that researchers will be able to
leverage the ISIS2015-2021 dataset to obtain a clearer understanding  of
how ISIS sympathizers use these platforms for message spreading and propaganda, especially in the recent years and with the advent of ISIS-K.  The dataset can also be used as a benchmark to investigate online strategies   originating from   extremist groups that are  not ISIS related. 
 In turn, such insight might enable  social network providers but also policy makers toward new interventions lessening the impact of extremist group influence.

\section{Ethical Considerations}
The dataset only collects data that is public and from alive accounts. Private accounts' tweet IDs and data are not included. Tweets limitations imposed by Twitter's API and ToS are respected. We have also removed all the tweet texts, full URLs, user account details, and only provide aggregated statistics, URL domains, and media keys that are not directly connected to any specific user or tweet and can be used for aggregate analysis to download the original tweets as long as they are available to the public. 

\bibliography{ref.bib}

\end{document}